\documentclass[12pt]{article}

\usepackage{mathptmx}       % selects Times Roman as basic font
\usepackage{helvet}         % selects Helvetica as sans-serif font
\usepackage{courier}        % selects Courier as typewriter font
\usepackage{type1cm}        % activate if the above 3 fonts are
 
\usepackage{graphicx}        % standard LaTeX graphics tool
\usepackage{multicol}        % used for the two-column index
\usepackage[bottom]{footmisc}% places footnotes at page bottom

\usepackage{amsmath}
\usepackage{amssymb}
\usepackage{amsfonts}
\usepackage{dsfont} % for unit operator  (indicator function) \mathds{1}

\newcommand{\be}{\begin{equation}}
\newcommand{\ee}{\end{equation}}
\newcommand{\bel}[1]{\begin{equation}\label{#1}}
\newcommand{\bea}{\begin{eqnarray}}
\newcommand{\eea}{\end{eqnarray}}
\newcommand{\ba}{\begin{array}}
\newcommand{\ea}{\end{array}}
\newcommand{\eref}[1]{(\ref{#1})}

\newcommand{\bra}[1]{\mbox{$\langle \, {#1}\, |$}}
\newcommand{\ket}[1]{\mbox{$| \, {#1}\, \rangle$}}
\newcommand{\exval}[1]{\mbox{$\langle \, {#1}\, \rangle$}}
\newcommand{\inprod}[2]{\mbox{$\langle \, {#1} \, | \, {#2} \, \rangle$}}
\newcommand{\Prob}[1]{\mbox{${\rm Prob}\left[ \, {#1}\, \right]$}}

\newcommand{\rmd}{\mathrm{d}}
\newcommand{\rme}{\mathrm{e}}
\newcommand{\erf}{\mathrm{erf}}
\newcommand{\bfr}{\mathbf{r}}
\newcommand{\bfn}{\mathbf{n}}

\newcommand{\T}{{\mathbb T}}
\newcommand{\Z}{{\mathbb Z}}
\newcommand{\C}{{\mathbb C}}
\newcommand{\R}{{\mathbb R}}

\newtheorem{theo}{Theorem}[section]

\newtheorem{prop}[theo]{Proposition}
\newtheorem{coro}[theo]{Corollary}

\def\qed{\hfill$\Box$\par\medskip\par\relax}

\begin{document}

\title{Conditioned stochastic particle systems and integrable quantum spin systems}
\author{Gunter M. Sch\"utz\\
\it \small Institute of Complex Systems II, Forschungszentrum J\"ulich, D-52425 J\"ulich, Germany \\
\it \small and
\it \small Fachgruppe Physik/Astronomie, Universit\"at Bonn\footnote{g.schuetz@fz-juelich.de}}

\maketitle

\abstract{We consider from a microscopic perspective large deviation properties of several stochastic interacting particle systems, 
using their mapping to integrable quantum spin systems. A brief review of recent work is given and
several new results are presented: (i) For the general disordered symmectric exclusion process (SEP) 
on some finite lattice conditioned on no jumps into some absorbing sublattice 
and with initial Bernoulli product measure with density $\rho$ we prove that the probability $S_\rho(t)$ of no absorption 
event up to microscopic time $t$ can be expressed in terms of the generating function for the particle number
of a SEP with particle injection and empty initial lattice. Specifically, for the symmectric simple exclusion process 
on $\Z$ conditioned on no jumps  into the origin we obtain the explicit first and second order expansion in $\rho$
of $S_\rho(t)$ and also  to first order in $\rho$ the optimal microscopic density profile under this conditioning. 
For the disordered ASEP on the finite torus conditioned on a very large current we show that the effective dynamics 
that optimally realizes this rare event does not depend on the disorder, except for the
time scale. For annihilating and coalescing random walkers we obtain the generating function of the 
number of annihilated particles up to time $t$,
which turns out to exhibit some universal features.
}

\section{Introduction}
\label{sec:1}

It is long known that the generator of some paradigmatic stochastic interacting particle systems such as the
asymmetric simple exclusion process (ASEP) \cite{Ligg99} on a one-dimensional lattice $\Lambda$ of $L$ sites
can be mapped to the Hamiltonian $H$ of an integrable quantum many-body system \cite{Alca93,Schu01}.
The non-diagonal matrix elements $h_{\eta'\eta} \leq 0$ of $H$ are the negative transition rates for a transition from
configuration $\eta$ to $\eta'$ of the particle system and each diagonal element $h_{\eta\eta} \geq 0$ is the
sum of transition rates out of a configuration $\eta$, i.e., the negative sum of the non-diagonal elements 
$h_{\eta'\eta}$ of column $\eta$. 
The interpretation as a quantum
many body system becomes evident by a suitable choice of tensor basis \cite{Schu01,Lloy96}. Consider
a particle system with local state space $S_i$ for each lattice site $i$ and and full state space $S=S_1\times \dots \times S_L$.
For definiteness we assume finite cardinality $0<|S_i|<\infty$ $\forall i$, even though much of what is discussed below 
can be generalized 
to infinite state spaces.
To each local configuration $\eta(i) \in S_i$ with $i\in\Lambda$ of the particle system 
one assigns a canonical basis vector $|\eta(i))$ of the vector space $\C^{|S_i|}$ and an associated dual basis vector
$(\eta(i)|$ with inner product $(\eta(i)|\eta'(i)) \delta_{\eta(i),\eta'(i)}$. The particle configuration $\eta = \{\eta(1),\eta(2),\dots,\eta(L)\} \in S$
on the lattice is then represented by the tensor vectors $\ket{\eta} = |\eta(1)) \otimes |\eta(2)) \otimes \dots \otimes |\eta(L))$
and their duals $\bra{\eta}$. A probability measure $P(\eta)$ is given by a probability vector $\ket{P}$
whose components are $P(\eta) = \inprod{\eta}{P}$. The semi-group property of the Markov process is reflected in
$\ket{P(t)} = \rme^{-H t}\ket{P(0)}$ for the time evolution of an initial measure $P_0(\eta)$. The expectation $\exval{f(t)}$ of a
random variable $f(\eta) $ is given by the scalar product $\exval{f(t)} = \bra{s} \hat{f} \ket{P(t)}$ where 
$\bra{s} = \sum_{\eta\in S} \bra{\eta}$ is called the summation vector and $\hat{f} = \sum_{\eta\in S} f(\eta) \ket{\eta}\bra{\eta}$
is a diagonal matrix (using the quantum mechanical shorthand $\ket{\cdot}\bra{\cdot} \equiv \ket{\cdot}\otimes \bra{\cdot}$)
with $f(\eta)$ on the diagonal. Because of probability conservation $\bra{s}$ is a left eigenvector of $H$ with eigenvalue 0.
Under this mapping the generator of the ASEP is represented by the quantum Hamiltonian
of a spin-1/2 Heisenberg ferromagnet with an Dzyaloshinskii--Moriya interaction with imaginary amplitude 
\cite{Gwa92,Kim95}, see next section.

Using this mapping one can employ rigorous methods borrowed from condensed matter theory 
to obtain information about the particle
system. However, due to the requirement of probability conservation and positivity of the classical transition rates
this mapping is restricted to quantum Hamiltonians with two special properties: 
\begin{itemize}
\item[(i)] All non-diagonal matrix elements of $H$ must satisfy $h_{\eta'\eta} \in (-\infty,0]$.\\

\item[(ii)] The lowest eigenvalue of $H$ is equal to zero for all lattice sizes $L$.
\end{itemize}
One can turn the mapping around and ask which integrable quantum Hamiltonian, or, in fact, which general finite-dimensional
matrix can be mapped to the generator of a Markov chain. 

The answer follows from the two properties (i) and (ii)
by using similarity transformations and the fact that adding a multiple of a unit matrix  to $H$ is a ``harmless'' operation in the
the time evolution operator $\rme^{-Ht}$: Imagine an irreducible square matrix $A$ which
has -- after a suitable similarity transformation -- the property $a_{\eta'\eta} \leq 0$ for all its off-diagonal elements 
$\eta\neq \eta'$.  Then we can
subtract a multiple $\lambda$ of the unit matrix $\mathds{1}$ such that $M = \rme^{-(A - \lambda \mathds{1})t}$
has only strictly positive matrix elements. It follows from Perron-Frobenius that the left largest eigenvector $\bra{0}$ of $M$ 
(and hence the left eigenvector to the 
lowest eigenvalue $a_0$ of $A$), has strictly positive components $\Delta(\eta)$. Next we define the diagonal matrix
$\Delta$ with these components on the diagonal. Since all components are strictly positive, its inverse
exists. Consider now $H := \Delta A \Delta^{-1} - a_0$. Since $\Delta$ is diagonal and $\Delta(\eta) > 0$ $\forall \eta$
we have that $h_{\eta'\eta} = a_{\eta'\eta}\Delta(\eta')/\Delta(\eta) \leq 0$. Moreover, the lowest eigenvalue of $H$ is 0
due to the subtraction of $a_0\mathds{1}$ from $\Delta A \Delta^{-1}$. Therefore $H$ is the transition matrix of some Markov chain.
This implies that an integrable quantum many-body system that satifies property (i) 
(perhaps after some similarity transformation), but {\it not} property (ii), can still be
mapped to a Markov chain which can then be analyzed using the mathematical machinery of 
quantum integrability.

There is a second reason to consider quantum Hamiltonians with property (i), but not (ii). This comes from the fact
that {\it conditioning} a process on some event may naturally lead to such a generator. In order to see this consider first 
conditioning a process on the event that a nonempty subset $X\subset S$ has never been reached until some
positive time $t$, when starting from an initial distribution with support in $A=S\setminus X$. 
We denote configurations in $X$ by $\xi$ and configurations in $A$ by $\alpha$.
This ``event conditioning'' can be 
studied by making the set $X$ absorbing and frozen,
i.e., all transition rates $w_{\alpha\xi}$ from $X$ to $A$ and transition rates $w_{\xi'\xi}$ within $X$ are defined to be zero.
The generator $H$ of such a process has a block tridiagonal structure 
$0_{|X|} \oplus G + H_{XA}$
with a zero-block $0_{|X|}$ of dimension $|X|$ (corresponding to the absence of transitions within $X$), 
a block $G$ of dimension $|A|$ containing all off-diagonal 
matrix elements $h_{\eta'\eta}$ for transitions within sector $A$ and
all diagonal elements for transitions out of  $A$, and the matrix $H_{XA}$ 
which has as only non-zero elements 
$h_{\xi\alpha}$, which are the negative transition rates from $A$ to $X$.
We also define a truncated basis where the basis vectors $\ket{\eta^A}$ ($\bra{\eta^A}$) are obtained from 
$\ket{\eta}$ ($\bra{\eta}$) (spanning the vector space $\C^{|S|}$) by
deleting the 0-components corresponding to configurations in $X$. The truncated canonical basis vectors
span $\C^{|A|}$. The truncated probability vector is denoted
by $\ket{P_0^A}$ and the truncated summation vector by $\bra{s^A} = \sum_{\eta\in A} \bra{\eta^A}$.

The block tridiagonal structure of $H$ has as a consequence that starting from an initial measure $P_0$
with support in $A$ the conditional probability of being in a state $\alpha \in A$ at time $t$, given that the process has 
never been in any configuration $\xi \in X$ until time $t$, is given by
\bel{1-1}
\tilde{P}(\eta,t) = \frac{1}{S_{P_0}(t)} \bra{\eta^A} \rme^{-Gt} \ket{P_0^A}
\ee
where
\bel{1-2}
S_{P_0}(t) = \bra{s^A} \rme^{-Gt} \ket{P_0^A}
\ee
is the survival probability of never having left $A$. By construction, $G$ satisfies
property (i), but not (ii). 
The matrix $\hat{H} = \Delta G \Delta^{-1} - g_0$ where $g_0$ is the lowest eigenvalue of $G$
and $\Delta$ the associated left eigenvector is then Doob's $h$-transform \cite{Doob}.
If $G$ is the quantum Hamiltonian of some integrable model then also $\hat{H}$ is integrable.

A different type of conditioning that has received strong interest in recent years arises
from incrementing some random variable $Q(t)$ by some value $c_{\eta'\eta}$
whenever a transition from $\eta$ to $\eta'$ occurs and
conditioning the process at time $t$ on reaching some fixed value $Q(t)=Q$.
Important applications include large deviations for the current or the activity in exclusion
processes \cite{Derr98,Leco12} (where the increments are integers) or the derivation of fluctuation
theorems \cite{Lebo99,Harr07}. In this paper  we restrict ourselves to integer increments.

In order to see the link to quantum Hamiltonians we go to a fluctuating ensemble
where instead of the joint probability $P(\eta,Q,t)$ one studies the generating
function $Y_(\eta,s,t) = \sum_Q \rme^{s Q} P(\eta,Q,t)$ with generalized
``chemical potential'' $s \in \R$. We refer to this conditioning at time $t$ 
as {\it grand canonical conditioning}.
One finds (see below) that
\bel{1-3}
Y(\eta,s,t) = \frac{1}{Y_s(t)} \bra{\eta} \rme^{-H(s) t} \ket{P_0}
\ee
with normalization factor
\bel{1-4}
Y_s(t) = \bra{s} \rme^{-H(s)t} \ket{P_0}
\ee
where $\bra{s} = \sum_{\eta\in S} \bra{\eta}$ is the summation vector and
the weighted generator $H(s)$ is obtained from $H$ by multiplying the off-diagonal elements
$h_{\eta'\eta}$ by a factor $\rme^{s c_{\eta'\eta}}$ where $c_{\eta'\eta}$ is the amount by which the counting variable
$Q$ has 
changed in the transition $\eta \to \eta'$. The diagonal elements of $H$ remain unchanged. Obviously, 
also $H(s)$ satisfies property (i), but not (ii). Notice that $H(0) = H$. 

In many cases of interest the integrability of the original Hamiltonian is not affected
by this modification. The matrix $H_{eff}(s) = \Delta(s) H(s) \Delta^{-1}(s) - \epsilon_0(s)$,
where $\epsilon_0(s)$ is the lowest eigenvector of $H(s)$
and $\Delta(s)$ the associated left eigenvector, is a generalized Doob's $h$-transform \cite{Jack10,Beli13a,Chet14}.
It defines a new process that we shall call {\it effective} process (or effective dynamics). For processes with
finite state space it can be interpreted as
that process which makes the untypical large deviation of $Q$ parametrized by $s$ typical \cite{Harr13}.
Evidently, if $H(s)$ is the quantum Hamiltonian of some integrable model then also the effective process is integrable.

We also define the conditioned history of some random variable $f$ which is the
expectation $\exval{f(t)}^c$ of $f$ at time $t$ of the process conditioned at time $T\geq t \geq 0$. For the event conditioning
this quantity is given by
\bel{1-5}
\exval{f(t)}^c = \bra{s_A} \rme^{-G(T-t)} \hat{f}  \rme^{-Gt} \ket{P_0}/S(t)
\ee
and for grand canonical conditioning by a similar expression with $G$ replaced by $H(s)$, summation vector
$\bra{s_A}$ replaced by the summation vector $\bra{s}$ over the full state space $S$ of the particle system
and $S(t)$ replaced by $Y_s(t)$.

For both types of conditioning we present some new results which are motivated by recent developments
in the field. For self-containedness and for reference we discuss the grand-canonical conditioning in more
detail and briefly review some results that we obtained earlier. Then we present our new results with a 
brief introduction into the history of the model that we treat.

\section{Conditioning on current and activity}

Following the seminal papers \cite{Derr02,Bert02}, the large deviation theory
for the ASEP has been developed in considerable detail, resulting in a powerful macroscopic
fluctuation theory (MFT) that is capable of describing large deviations of the current and density profiles
of quite general lattice gases on a macroscopic level, see \cite{Bert14} for a recent review. 
Here the notion of ``macroscopic'' refers to the scale where the lattice constant $a$ is taken to
zero and the lattice site $k$ is taken to infinity with finite rescaled macroscopic space variable $x=ka$.
The microscopic time is taken to infinity under  rescaling by a factor $1/a^z$ where $z=1$ 
for Eulerian scaling and $z=2$ for diffusive scaling. The {\it typical} macroscopic density profile of the ASEP, i.e.,
without conditioning on some large deviation, 
is obtained from the usual hydrodynamic limit \cite{Reza91}.
%\footnote{We shall use the symbol $t$ both for
%microscopic time and rescaled macroscopic time, as it will always be clear from context which is
%meant. Obviously, the macroscopic description also involves the limit $L\to\infty$ so that a periodic lattice
%of $L$ sites becomes a ring of macroscopic length 1. This limit will not be used explicitly.}

However, large deviations can also be studied directly on a microscopic level. To illustrate this
we focus on large deviations of the integrated hopping activity in the asymmetric simple exclusion process
(ASEP) on the torus $\T_L:=\Z/L\Z$. Particles jump according to the following rules:
\begin{enumerate}
\item Each particle attempts to jump independently of the other particles
after an exponentially distributed random time with parameter $1/(w_+ + w_-)$ 
with probability  $w_+/(w_+ + w_-)$ to the next site on the right (clockwise) and $w_-/(w_+ + w_-)$ 
to the next site on the left (counterclockwise)
\item The hopping attempt is rejected if the
site to which the particle tries to move is occupied.
\end{enumerate}
We introduce the hopping asymmetry 
$q^2:=w_+/w_-$ and time scale $w_0 = \sqrt{w_+ w_-}$ and
assume without loss of generality a hopping bias to the right, i.e., $q>1$.

The local state space of the ASEP is $S_i=\{0,1\}$ $\forall i \in \T_L$,
corresponding to a vacant or occupied site $i$ with occupation variable $n_i=0$ or $n_i=1$.
The full state space therefore is therefore $S=\{0,1\}^L$ with microscopic configurations
$\bfn = \{n_1,\dots,n_L\}$.
Following the quantum Hamiltonian formalism outlined above we introduce the usual Pauli matrices $\sigma^{x,y,z}$ 
and the two-dimensional unit matrix $\mathds{1}$. From these we construct
operators $a_k$ acting locally on site $k$
of the lattice in terms by the tensor product
\be
\mathds{1} \otimes \dots \otimes a \otimes \dots \mathds{1}
\ee
with the arbitrary $2\times 2$ matrix $a$ at position $k$ in the product.
This allows us to define local
particle creation and annihilation operators 
$\sigma^\pm_k = (\sigma^x_k \pm  i \sigma^y_k)$  and projectors 
$\hat{n}_k = (\mathbf{1} - \sigma^z_k)/2$
on particles and $\hat{v}_k = \mathbf{1}-\hat{n}_k$ on vacancies on site $k$ where $\mathbf{1} = \mathds{1}^{\otimes L}$ is the
unit matrix of dimension $2^L$. Here the local basis vector is chosen such that
$|0) = (1,0)^T$ and $|1) = (0,1)^T$ where the superscript $T$ denotes transposition. Correspondingly
$(0| = (1,0)$ and $(1|=(0,1)$ for the dual basis. A state $\bfn \in S$ is then represented by
$\ket{\bfn} = |n_1)\otimes \dots \otimes |n_L)$ and
the generator is given by \cite{Schu01,Gwa92}

\be
\label{2-1-1}
H = - \sum_{k=1}^L 
w_+ (\sigma^+_k \sigma^-_{k+1} - \hat{n}_k \hat{v}_{k+1})+
w_- (\sigma^-_k \sigma^+_{k+1} - \hat{v}_{k} \hat{n}_{k+1} )
\ee

with the identification of site $L+1$ with site $1$. 
The total particle number $N$ is conserved, which
is expressed in the commutation relation $[H,\hat{N}]=0$ for the number operator $\hat{N} = \sum_{k=1}^L
\hat{n}_k$. 
It is well-known that \eref{2-1-1} defines an integrable model related to the six-vertex model \cite{Gwa92,Kim95}.

The unique stationary distribution of the process with $N$ particles is the uniform measure with
$P^\ast(\bfn)= 1/{L \choose N}$. We refer to this invariant measure as canonical. From these measures one can
construct the grand-canonical Bernoulli product measure with $P^\ast(\bfn)= \rho^{N(\bfn)}(1-\rho)^{L-N(\bfn)}$ where
$N(\bfn) = \sum_{k=1}^L n_k$. The grandcanonical measure has the vector representation
$\ket{\rho} := ((1-\rho,\rho)^T)^{\otimes L}$, reflecting its product structure. In the thermodynamic
limit $L \to \infty$ with $\rho=N/L$ fixed the canonical measure and its grand canonical counterpart
become equivalent. The stationary current in the grand canonical measure is given by
$j^\ast = (w_+-w_-) \rho(1-\rho)$.

Now we consider the time-integrated number of jumps $J_k^+(t)$ from site $k$ to $k+1$ up to time $t$,
and analogously the time-integrated number of jumps $J_k^-(t)$ from site $k+1$ to $k$ up to time $t$.
The time-integrated local current is then given by  $J_k(t)=J_k^+(t) - J_k^-(t)$ and $A_k(t)=J_k^+(t) + J_k^-(t)$
is the time-integrated local activity and $j_k(t) := J_k(t)/t$,  $a_k(t) := A_k(t)/t$ the respective
time averages. We also define the space-integrated quantities $J^{\pm}(t) = \sum_{k=1}^L J^{pm}_k(t)$
and analogously $J(t)$ and $A(t)$ as well as their space-time averages $j^{\pm}(t) = J^{\pm}(t)/(Lt)$ and
analogously $j(t)$ and $a(t)$. 
%Since we have a finite local state space, the local quantities $J_k^{\pm}(t)$
%differ from $J^{\pm}(t)$ by an amount that is bounded in  $L$. 

The probability to observe for a long time interval
$t$ an untypical mean $j\neq j^\ast$ is exponentially small in $L$ and $t$. 
This is expressed in the asymptotic large deviation property
$\Prob{J(t) = J} \propto \exp{(-f(j)Lt)}$ \cite{Derr98}  where $f(j)$ is the rate
function and one takes the limit $L\to\infty$ and $t\to\infty$. It is then natural to
introduce a generalized fugacity $y=\rme^s$ with generalized
chemical potential $s$ and to study the generating
function $Y^c_s(t) := \sum_{J\in\Z} y^J \Prob{J(t) = J}$. 
The cumulant function $g(s) = \lim_{t\to\infty} \ln{Y^c_s(t)}/(Lt)$ is the 
Legendre transform of the rate function, i.e., $g^c(s) = \max_j [ j s - f(j)] $.
The intensive variable $s$ is thus conjugate to the mean current density $j$. 
Similar large deviation properties hold for the activity $a(t)$ with generating function
$Y^a_s(t)$ and the hopping activities $j^\pm(t)$ with generating functions
$Y^\pm_s(t)$ and their local counterparts $Y^{\cdot}_s(k,t)$.

Following \cite{Derr98,Harr07} one has for an initial distribution $P_0$ of particles
\be
Y^{\cdot}_s(k,t) = \bra{s} \rme^{-H^{\cdot}(k,s) t} \ket{P_0}
\ee
where $H^{\cdot}(k,s)=H+F^{\cdot}(k,s)$ with
\be
F^{+}(k,s) = (1-\rme^s) w_+ \sigma^+_k \sigma^-_{k+1}, \quad
F^{-}(k,s) = (1-\rme^s)  w_- \sigma^-_k \sigma^+_{k+1}
\ee
and $F^{c}(k,s) = F^{+}(k,s) - F^{-}(k,-s)$ and $F^{a}(k,s) = F^{+}(k,s) + F^{-}(k,s)$.
These are examples for the general construction \eref{1-3}, \eref{1-4}, detailed in \cite{Harr07}.
For the ASEP and other integrable models $g^{\cdot}(s)$ and other large deviation functions,
e.g. associated with particle creation and annihilation events, have been computed with the
Bethe ansatz \cite{Derr98,Derr99,Appe08,Simo09,Cast14} or free fermion methods \cite{Popk11b}.

The MFT provides also information about the optimal macroscopic density profile $q(x,t)$
to realize at time $t$
a prolonged untypical behaviour of the integrated current at time $T\geq t$, i.e., the density profile that one is
most likely to observe at time $t$ in the interval $[0,T]$.
One expects the optimal macroscopic density profile to be the large scale limit of the
conditional history \eref{1-5} of the local particle number.
To our knowledge no  microscopic computation of this large deviation property has been performed
yet.

Using MFT for the
weakly asymmetric exclusion process (WASEP), where $ w_+-w_- = \nu/L$ is small,
Bodineau and Derrida \cite{Bodi05} have considered the time range $t=cT$ with $c\in(0,1)$.
In the limit $T\infty$ and $L\to\infty$ they found a dynamical phase transition 
at some critical $s_c < 0$ (i.e., for conditioning on untypically low currents $j<j^\ast$), where
the optimal macroscopic density
profile $q(x,t)$ changes from a constant $q(x,t)=\rho$ to a travelling wave
of the form $q(x,t)=\rho(x-vt)$. For
strictly positive $s$ (i.e., for any current $j$ 
conditioned on $j>j^\ast$) 
is time-independent and flat and hence equal to the typical profile $\rho(x,t)= \rho$.

The macroscopic approach is extremely powerful, but has the drawback that no information is available
on the expected non-trivial space-time realization of the ASEP  conditioned on currents larger than
the typical value $j^\ast$. Also, no information can be obtained about fluctuations of the traveling wave
for $s<0$. This gap was partially filled in a series of papers \cite{Beli13a,Simo09,Popk10,Popk11a,Schu14},
exploiting the integrability and symmetries of the weighted generator $H(s)$.

Significantly, this microscopic approach shows that for conditioning on high currents
or high activities (more precisely, in the limit
$s\to\infty$ on the time scale set by $w_0=\rme^{-s} q^{-1}$)
there is a different dynamical phase transition that does not appear in the density profile,
but in the fluctuations. The dynamical exponent for density fluctuations changes from $z=3/2$ for typical behaviour
to $z=1$ for atypical behaviour for times $t=cT$ with $c\in(0,1)$ and $T\to\infty$.
The associated effective microscopic dynamics, obtained from the generalized Doob's transform, 
have nearest neighbour jumps, but the jump rate depends
on the whole configuration of the particles in the same way as in a process introduced by Spohn
\cite{Spoh99}. The stationary distribution of the effective process with $N$ particles
located on sites $k_1,\dots,k_N  \in \T_L$ has the determinantal form
\begin{equation}
\label{2-2}
P_{L}(\bfn)=
\frac{2^{N(N-1)}}{L^{N}} \prod\limits_{1\leq i<j\leq N} \sin^{2}
\left(\pi \frac{k_{i}-k_{j}}{L} \right)
\end{equation}
with long-range anticorrelations in the thermodynamic limit $L\to\infty$
\begin{equation}
\label{2-3}
C(k) := \exval{n(0) n(k)} - \rho^2 = -\frac{\sin^{2}(k\pi\rho)}{k^{2}\pi^{2}}.
\end{equation}

For a current lower than typical we considered in \cite{Beli13a} the grand canonically conditioned
ASEP at $t=T$ finite for a special
value of the conjugate variable $s$ and a special family of initial measures that are inhomogeneous
product measures mimicking a microscopically sharp shock and some site $k$ and a soft antishock extending over the
whole lattice $\T_L$. It turns out that under the
conditioned dynamics this measure evolves into a time-dependent convex combination of such measures
with microscopic shock positions at sites $l$ where the weights for each such measure are the transition
probabilities of a biased simple random walk starting from site $k$ at $t=0$ and ending at site $l$ at time $t=T$
and explicitly known random walk transition rates \cite{Beli13a}.
Thus under these conditions one has on macroscopic level a travelling wave with shock moving around the
ring with finite speed, reminiscent of what is predicted by MFT at times $cT$ with $c\neq 1$. 
In addition to this macroscopic information one finds that the shock position
performs diffusive fluctuations around its mean position with an explicitly computed shock diffusion coefficient. 
Moreover, the shock remains microscopically
sharp at all times and there are no correlations seen from the shock position.
The techniques used to derive these results are an adaptation of our earlier work \cite{Beli02} on the microscopic structure of
shocks under unconditioned dynamics. This approach, which uses the $q$-deformed
non-abelian symmetries of the integrable weighted generator, has a probabilistic
interpretation as self-duality \cite{Schu87,Giar09,Imam11,Boro12}.

\section{New applications of conditioning}

After this brief review of microscopic large deviation results we consider now some models that have a 
long history of study, but where there is renewed interest from the viewpoint of large deviations.

\subsection{SEP conditioned on not entering a sublattice}

Recently Meerson et al. \cite{Meer14} studied absorption into a static target in a gas of diffusing particles with exclusion,
using macroscopic fluctuation theory (MFT) for the symmetric simple exclusion process (SSEP). 
The target is considered to be an absorber, i.e., when a particles hits the target it
disappears from the dynamics. A quantity Meerson and collaborators are interested in is the probability $S(t)$ that no gas particle
has hit the target until time $t$. This conditioning on an event is of the first type outlined in the introduction. For non-interacting particles
this problem has appeared recently also in the search for an immobile target by a swarm of diffusive searchers \cite{Meji11}
and in the study of the complete statistics of particle absorption by the absorber \cite{Redn14}.

For an initial distribution  that is a Bernoulli
product measure with density $\rho$ on all sites not covered by the target, MFT predicts the large deviation form \cite{Meer14}
\bel{3-1-0a}
S(t) \propto \rme^{-f(\rho)\sqrt{t}}
\ee
with a large deviation function $f(\rho)$. 
In one dimension, by placing the right edge of the target at site 0, this problem is equivalent
to conditioning the SSEP on the event that no particle has reached site 0 until time $t$. 
For this case Meerson et al. computed from MFT explicitly the first two terms in the Taylor expansion
of \eref{3-1-0a}, with the result
\bel{3-1-0b}
f(\rho) = \frac{2}{\sqrt{\pi}} \left[ \rho + (\sqrt{2}-1) \rho^2 + \dots \right]
\ee
where the dots represent terms of higher order in $\rho$. The first order term, corresponding to non-interacting particles, has been known for a long time,
see \cite{Meer14} and references therein.

Here we take a microscopic approach and first generalize to the disordered symmetric exclusion
process on any finite lattice $\Xi$, i.e., a lattice where hopping is allowed with strictly positive rate 
between fixed, but arbitrary pairs 
$<i\,,j>$ of sites with $i,j\in\Xi$ and such that the hopping rates may depend in the link 
$<i\,,j>$.\footnote{To clarify our terminology: The symmetric simple
exclusion process (SSEP) has nearest neighbour jumps only on a regular lattice such as $\Z^d$, whereas the general symmetric 
exclusion process (SEP) can have jumps between any pair of nodes on any graph. This terminology comes from physical intuition
and is not mathematically precise, since the set of links on which jumps have non-zero rate may just as well
be used to define an underlying graph on which one then would again have only nearest neighbour jumps
so that the process could also be called SSEP.
However, we will only consider the SSEP on $\Z$ with bond-independent rates. This removes any ambiguity in our results.
Moreover, we will always speak of sites and lattices rather than nodes and graphs.}
We show for arbitrary targets that the conditioned process is equivalent to an unconditioned SEP with particle injection.
Then we focus on the case of Meerson et al. For the SSEP in one dimension the equivalent process
was studied earlier \cite{Sant01}. Using results from that work
allows us to obtain the fluctuations
of the conditioned process in a fully microscopic approach from the Bethe ansatz and to prove the 
(not fully rigorous) prediction \eref{3-1-0b} of MFT.

According to the construction outlined above, in one dimension (where $\Xi = \Z$) 
the SSEP conditioned on no particle reaching the origin
is given by the truncated generator 
\bel{3-1-1a}
G_{\Z^+} = - w \sum_{k=1}^\infty \left[ \sigma_k^+ \sigma_{k+1}^- + \sigma_k^- \sigma_{k+1}^+ - \hat{n}_k \hat{v}_{k+1}
- \hat{v}_k \hat{n}_{k+1} \right] + w \hat{n}_1 =: G_0 + w \hat{n}_1
\ee

For general finite lattices $\Xi$ and general targets, covering an arbitray (but non-empty) sublattice
$\Omega \subset \Xi$,
and disordered symmetric hopping with exclusion between any pair of sites we by denote by $\Lambda \subset \Xi$
the sublattice of sites not covered by the target and between which hopping has non-zero rate.
Hopping across a bond $<i\,,j>$ with $i,j \in \Lambda$ occurs symmetrically
with bond-dependent rate $w_{i,j}=w_{j,i}$.
Furthermore we denote by ${\mathcal E} \subset \Lambda$ the exit sites which allow for hopping into the 
target in the unconditioned process. (For nearest neighbour hopping  e.g. on a cubic lattice these sites form the 
external boundary of the target.) 

Then one has as similar truncated generator $G_{\Lambda}$ with projectors
$\hat{n}_k$ on the set of sites $k \in {\mathcal E}$
from which jumps into the target can occur, viz.,
\bel{3-1-1b}
G_{\Lambda} = - \sum_{<i,j>} w_{i,j} \left( \sigma_i^+ \sigma_{j}^- + \sigma_i^- \sigma_{j}^+ - \hat{n}_i \hat{v}_{j}
- \hat{v}_i \hat{n}_{j} \right)  + \sum_{i \in {\mathcal E}} d_i \hat{n}_i =: G_{\Lambda}^0 +G_{\Lambda}^e
\ee
where $d_i = \sum_{j\in \Omega} w_{i,j}$ is the sum of jump rates between $i$ 
and the target region to which a particle in the unconditioned process can jump from site $i$.
The survival probability $S_{P_0}(t)$ is then given by
\bel{3-1-2}
S_{P_0}(t) = \bra{s} \rme^{-G_{\Lambda}t} \ket{P_0}
\ee
where $\bra{s}$ and the initial distribution $\ket{P_0}$ are restricted to configurations with support on the sites 
in $\Lambda$ outside the target, i.e., to all configurations with $n_l=0$ for $l \in \Omega$.

Our main result is the following.
\begin{theo}
\label{SEPcond}
Let $\Xi$ be a finite lattice with $M$ sites, $\Omega$ be the sublattice with $K$ sites covered by fixed
absorbers and $\Lambda = \Xi \setminus \Omega$ with $L = M - K$ sites. Then:\\

\noindent (a) For initial product measure with
density $\rho$ one has
\bel{3-1-3}
S_\rho(t) = \exval{(1-\rho)^N}_0
\ee
where $N$ is the total particle number under the evolution of an unconditioned SEP 
with injection of particles at the sites $i \in {\mathcal E}$
with rate $d_i= \sum_{j\in \Omega} w_{i,j}$, defined by the generator
\bel{3-1-4}
H_{\Lambda} = -  \sum_{<i,j>} w_{i,j} \left( \sigma_i^+ \sigma_{j}^- + \sigma_i^- \sigma_{j}^+ - \hat{n}_i \hat{v}_{j}
- \hat{v}_i \hat{n}_{j} \right)  -  \sum_{i \in {\mathcal E}} d_i (\sigma^-_i - \hat{v}_i) =: H_{\Lambda}^0 +H_{\Lambda}^e
\ee
and $\exval{\cdot}_0$ represents expectation for the initial distribution representing the empty lattice.\\

\noindent (b) For initially $N$ particles placed with probability 1 on sites $k_1,\dots,k_N \in \Lambda$ one has
\bel{3-1-5}
S_{k_1,\dots,k_N}(t) = \exval{v_{k_1}\dots v_{k_N}}_0
\ee
for the process with open boundaries defined by \eref{3-1-4} and initially empty lattice.
\end{theo}

{\it Proof:}
Observe that $\bra{s} = \bra{0} U$ where $U = (\hat{v} + \sigma^x)^{\otimes L}$ and that 
$G_{\Lambda}^0=H_{\Lambda}^0$ is symmetric
under the action of the generators $S^{x,y,z}=\sum_{k\in\Lambda}\sigma^{x,y,z}_k$ of the Lie-group $SU(2)$ 
and therefore commutes
with $U$ \cite{Schu94}. Moreover, by explicit computation 
$U \hat{n}_i U^{-1} = \hat{v}_i - \sigma_i^+$. These relations are valid for any lattice. 
Therefore $\tilde{G}_{\Lambda} := UG_{\Lambda}U^{-1} = G_{\Lambda}^0 -  \sum_{i\in {\mathcal E}}d_i(\sigma_i^+-\hat{v}_i)$ 
which yields for an initial product measure
$S_\rho(t) = \bra{0} \rme^{-\tilde{G}_{\Lambda}t} U \ket{\rho}$. Next we observe that  $U \ket{\rho} = (1-\rho)^{\hat{N}} \ket{s}$
and $H_{\Lambda} = \tilde{G}_{\Lambda}^T$
where the superscript $T$ denotes transposition.
Transposition of \eref{3-1-2} then proves \eref{3-1-3}. On the other hand, for fixed initial condition with $N$ particles on
sites $k_1,\dots,k_N$ one has $U \ket{k_1,\dots,k_N} = \hat{v}_{k_1} \cdots \hat{v}_{k_N}  \ket{s}$ 
since $ (\hat{v} + \sigma^x) |0)= |s)$ and
$ (\hat{v} + \sigma^x) |1)= |0) = \hat{v} |s)$. Transposition of \eref{3-1-2} then yields \eref{3-1-5}.
\qed

In order to avoid irrelevant technicalities regarding arbitrary infinite lattices we have restricted ourselves to
finite lattices $\Xi$. However, all the steps involved in the previous proof remain valid for the SSEP on $\Xi=\Z^d$. In particular,
for the SSEP on $\Xi=\Z$ with constant jump rates $w$ and right target edge at site 0 we have $\Lambda=\Z^+$.
The process \eref{3-1-4}
derived from $G_{\Z^+}$
corresponds to injection of particles at site 1 with rate $w$ with generator
\bel{3-1-4c}
H = - w \sum_{k=1}^\infty \left[ \sigma_k^+ \sigma_{k+1}^- + \sigma_k^- \sigma_{k+1}^+ - \hat{n}_k \hat{v}_{k+1}
- \hat{v}_k \hat{n}_{k+1} \right] - w  (\sigma^-_1 - \hat{v}_1)
\ee
and initial density $\rho_0=0$. This is the special case $\rho^\ast=1$,
$\rho_0=0$ of the SSEP with open boundaries where particles are injected at site 1 with rate $w\rho^\ast$ and extracted
with rate $w(1-\rho^\ast)$ and the lattice has an initial product distribution with density $\rho_0$, see \cite{Sant01} with the
(arbitrary) choice of time scale $w=1$.

\begin{coro}
From \eref{3-1-3} it follows that the lowest order terms in the Taylor expansion of $\ln{S_\rho(t)}$ in $\rho$ are given by 
\be
\ln{S_\rho(t)} = -\exval{N}_0 \rho + \frac{1}{2}\left( \exval{N(N-1)}_0-\exval{N}_0^2 \right) \rho^2 + \dots
\ee
\end{coro}

Using rigorous Bethe ansatz techniques these expansion coefficients have been computed 
in \cite{Sant01} for arbitrary boundary density $\rho^\ast$ 
and arbitrary initial density $\rho_0$. They  agree for the relevant case $\rho^\ast=1$, $\rho_0=0$ with the
MFT prediction \eref{3-1-0b} which is thus proved.

Next we discuss the conditioned time evolution of the local density in the general SEP.
\begin{prop}
For the SEP \eref{3-1-1b} conditioned on no absorption in the
target up to time $T\geq t \geq 0$ let 
\be
\phi_{\rho,T}(k,t) = \frac{\bra{s} \rme^{-G_\Lambda (T-t)} \hat{n}_k \rme^{-G_\Lambda t} \ket{\rho} }{S_\rho(T)}
\ee
be the expected intermediate local particle number with initial Bernoulli product measure with density $\rho$.  

(a) This function has  $\forall \rho \in [0,1]$  and $\forall t \in [0,T]$ the time reversal symmetry
\bel{3-1-6}
\phi_{\rho,T}(k,t) = \phi_{\rho,T}(k,T-t)
\ee
(b) and the property
\bel{3-1-7}
\phi_{\rho,T}(k,t) = \frac{\bra{s} (1-\rho)^{\hat{N}} \rme^{-H_\Lambda t} (\hat{v}_k - \sigma_k^-) \rme^{-H_\Lambda (T-t)} \ket{0} }{S_\rho(T)}.
\ee
\end{prop}

{\it Proof:}
(a) The first statement \eref{3-1-6} follows from a reversibility argument: One uses particle number conservation 
$G_\Lambda = y^{\hat{N}} G_\Lambda y^{-\hat{N}} = y^{\hat{V}} G_\Lambda y^{-\hat{V}}$, $\forall y\in\C$,
and the time reversal symmetry $G_\Lambda=G_\Lambda^T$ of the conditioned evolution and notes that 
$\ket{\rho} = (1-\rho)^{\hat{V}} \rho^{\hat{N}} \ket{s}$. (b) The second equality \eref{3-1-7} is obtained using the 
same transformations as in the proof of \eref{SEPcond}. \qed

For $\Lambda=Z^+$ \eref{3-1-7} can be evaluated to first order in $\rho$ using symmetries and elementary 
computations. We define
\be
\label{3-1-8}
q_{T}(k,t)  :=  - \rho \frac{\rmd }{\rmd \rho}\big|_{\rho=0}  \phi_{\rho,T}(k,t)
\ee
and obtain
\be
q_{T}(k,t) = \rho \sum_{l=1}^\infty \bra{s} \hat{n}_l \rme^{-H t} (\hat{v}_k - \sigma_k^-) \rme^{-H (T-t)} \ket{0}
\ee
with  $H$ given by \eref{3-1-4c}. This brings us into a position to state:

\begin{prop}
Let $q_{T}(k,t)$ be the conditional expectated particle number at site $k$ to first order in the initial density $\rho$
of the SSEP on $\Z$, conditioned on the event that no particle has crossed the origin up to time $T\geq t \geq 0$, with initial
product measure with density $\rho$ on $\Z^+$. One has
\bel{3-1-9}
q_{T}(k,t) = \rho f_k(2wt) f_k(2w(T-t))
\ee
where
\be
f_k(u) = \rme^{-u} \sum_{n=-k+1}^k I_n(u)
\ee
with the modified Bessel function $I_n(\cdot)$. 
\end{prop}

{\it Proof:}
In order to prove this result one first observes that
$\bra{s} \hat{n}_m \rme^{-H t} (\hat{v}_k - \sigma_k^-) = - \bra{s} \hat{v}_m \rme^{-H t} (\hat{v}_k - \sigma_k^-)$.
Next one shows by direct computation that the vector $\bra{m,t}:= \bra{s} \hat{v}_m \rme^{-H t}$ satisfies the
lattice diffusion equation $\rmd/(\rmd t) \bra{m,t} = w  (\bra{m+1,t} + \bra{m-1,t} - 2 \bra{m,t})$ with absorbing
boundary condition $\bra{0,t}=0$ $\forall t$. Hence $\bra{m,t} = \rme^{-2wt} \sum_{n=-m+1}^m I_n(2wt) \bra{n,0}$
which yields 
\be
q_{T}(k,t) =  \rho\, \rme^{-2wt} \sum_{n=-m+1}^m I_n(2wt) \bra{s} \hat{v}_n (\hat{v}_k - \sigma_k^-) \rme^{-H (T-t)} \ket{0}.
\ee
Now observe that $\bra{s} \hat{v}_n (\hat{v}_k - \sigma_k^-) = \bra{s} \hat{v}_n \delta_{n,k}$ and therefore one has
$q_{T}(k,t) =  \rho f_k(2wt) \bra{s} \hat{v}_k  \rme^{-H (T-t)} \ket{0}$. Now one repeats the previous computation 
for the time evolution of $\bra{s} \hat{v}_k$ (with $t$ replaced by $T-t$)
and uses $\bra{s} \hat{v}_n   \ket{0}=1$. This yields \eref{3-1-9}. \qed

At large scales where $u$ is large and $k = O(\sqrt{u})$
the function $f_k(u)$ converges to the error function. We rescale time $t=\tau T$, with $0 \leq \tau \leq 1$,
and space $k = x \sqrt{wT}$.
Then \eref{3-1-9} becomes
\bel{3-1-10}
\lim_{T\to\infty} q_{T}(k,t) = \rho\, \erf{ \left(\frac{x}{2\sqrt{\tau}}\right)} \erf{\left(\frac{x}{2\sqrt{1-\tau}}\right)}.
\ee
A similar result was obtained from MFT for the macroscopic optimal density profile
of non-interacting random walks \cite{Meer14}
conditioned on no absorption at the origin.

\subsection{ASEP with a defect bond conditioned on high current}

The TASEP on $\T_L$ with a defect bond is the TASEP essentially as described
above, with the difference that hopping attempts across bond $<L\,,1>$ happen with rate $r$ and $0<r<1$.
This process has a long history of study which started with the numerical observation of a 
defect-induced phase transition first in a closely related model of interface growth \cite{Wolf90} and then with a
numerical study by Janowsky and Lebowitz \cite{Jano92} who also reported a defect-induced phase transition 
from a phase with homogenous density (on coarse-grained scale) to a phase-separated stationary state with two domains
of different densities. A major issue that is still open from a mathematical perspective is whether the phase
 transition occurs at some finite defect strength $r_c$ or for arbitrarily small defects, i.e., at $r=1$. 
In a related simpler TASEP with
discrete-time update
\cite{Schu93a} this question could be answered rigorously. In that model the phase transition sets in for any defect
strength. Tang and Lyuksyutov \cite{Tang93} predicted this behaviour also for the original TASEP, using renormalization group
arguments for the closely related directed polymer problem. For more recent work on this
problem, both in mathematical probability theory and statistical physics, see \cite{Baha04,Szav09} and references therein.

Here we follow the line of investigation started in \cite{Popk10}
and consider the ASEP with a defect conditioned on a large current, parametrized by the conjugate
variable $s$. Hopping attempts across bond $<L,1>$
happen with rate $r w_{\pm}$ and $0<r<1$. Following the construction outlined above
one has to study in the maximal current limit $s\to\infty$ the rescaled weighted generator
\bel{3-3-1}
H^\ast(r) = - \sum_{k=1}^{L-1} \sigma_k^+\sigma_{k+1}^- - r  \sigma_k^+\sigma_{k+1}^-
\ee
Solving for the conditioned stationary distribution and the conditioned dynamics looks at first sight
like a difficult spatially inhomogenous many-body problem. However, the inhomogeneity turns out to 
have an trivial effect. In order to see this we prove 

\begin{prop}
The weighted generator \eref{3-3-1} with bulk hopping rate 1 and defect hopping rate $r=\rme^{-u}$ is related to a 
homogeneous generator with hopping rate $c(r) = r^{1/L} = \rme^{-u/L}$ by the similarity transformation
\bel{3-3-2}
V_r H^\ast(r) V_r^{-1} = c(r) H^\ast(0).
\ee
with $V_r = \rme^{-\frac{u}{L} \sum_{k=1}^L k \hat{n}_k}$.
\end{prop}

{\it Proof:}
This is proved by 
straightforward computation, using the factorized form $V_r = \prod_{k=1}^L V_r(k)$ with
$V_r(k) =  \rme^{-\frac{u}{L} k \hat{n}_k}$ and the transformation
property $\rme^{-\alpha \hat{n}} \sigma^\pm \rme^{\alpha \hat{n}} = \rme^{\pm \alpha} \sigma^\pm$ of the Pauli matrices. \qed

An obvious corollary is that all eigenvalues $\epsilon_i(r)$ of the weighted generator \eref{3-3-1} are given by 
\bel{3-3-3}
\epsilon_i(r) = c(r) \epsilon_i(1)
\ee
in terms of the eigenvalues of the homogeneous generator. For the left and right eigenvectors to these eigenvalues one has
$\bra{\epsilon_i(r)} = \bra{\epsilon_i(1)} V_r $ and $\ket{\epsilon_i(r)} = V_r^{-1} \ket{\epsilon_i(1)}$. The eigenvalues
and eigenvectors for $r=1$ are computed in \cite{Popk10} using the integrability of the model by free fermion techniques.

Another immediate consequence is stated in the following theorem:

\begin{theo}
\label{inhomoASEP}
The effective dynamics for conditioning on maximal current in the disordered ASEP given by 
the generalized Doob-transform $H_{eff}^\ast(r) := \Delta_r H^\ast(r) \Delta_r^{-1} - \epsilon_0(r)$ of \eref{3-3-1} 
has the property
\bel{3-3-4}
H_{eff}^*(r) = c(r) H_{eff}^*(1).
\ee
\end{theo}

This follows from $\Delta_r = \Delta V_r$ with $\Delta = \Delta(1)$ given in \cite{Popk10}.
The interpretation of this at first sight surprising fact is the following. The conditioning on maximal current
selects those (untypical) realizations of the process where hoppping attempts occur instantly, irrespective of the
actual rate (which determines the typical mean frequency between hopping attempts). Only the time
scale of the effective process (rescaled by $\rme^s$) is affected by the defect. As an obvious but noteworthy 
corollary we find that the
stationary distribution of the transformed process \eref{3-3-4} does not depend on $r$ and is given by \eref{2-3}.

This result can be extended to the fully disordered ASEP, which also has a long history of study, see
\cite{Szav11} and references therein. We denote the set of site-dependent hopping rates $r_k=\rme^{-u_k}$ to the right 
by $\bfr$ and obtain under conditioning on maximal current
\bel{3-3-6}
H^*(\bfr) = - \sum_{k=1}^{L} r_k \sigma_k^+\sigma_{k+1}^- .
\ee
with the property
\bel{3-3-7}
V_{\bfr} H^\ast(\{r\}) V_{\bfr}^{-1} = c(\bfr) H^\ast(0).
\ee
where
\bel{3-3-8}
V_{\bfr} =\rme^{- \sum_{k=1}^L \alpha_k \hat{n}_k}, \quad c(\bfr) = \rme^{-\frac{1}{L} \sum_{k=1}^L u_k}
=\left( \prod_{k=1}^L r_k \right)^{\frac{1}{L}}
\ee
and $\alpha_k = - \sum_{i=1}^{k-1} (u_k - c(\bfr))$. Theorem \eref{inhomoASEP} applies also to this case
with the obvious substitution $r\to\bfr$. Hence hopping disorder is irrelevant in the ASEP under maximal current
conditions.

\subsection{Annihilating and coalescing random walks}

We consider non-conservative dynamics where random walks
annihilate instantly when they meet, also called vicious random walkers \cite{Fish84}. There is a vast body of literature, both 
in statistical physics and probability theory, on this
process and the related process of coalescing random walks where only one particle annihilates upon meeting,
Many different techniques (see e.g. \cite{Krap10,Colo93,Schu95,Ayye10,Kato13} and references therein) allow
for a detailed analysis of this problem.
It is known for a long time that on $\Z$ the average particle density approaches 0 proportional to $1/\sqrt{t}$,
irrespective of the initial density (after some finite crossover time) \cite{Bram80}. In the spirit of \cite{Popk11b} 
we study here for vicious walkers on $\T_L$ the annihilation activity $A^{ann}(t)$, i.e., the number of particles annihilated until time $t$. 
Thus we are dealing with the weighted generator
\bel{3-4-1}
H^a(s) = - \sum_{k=1}^L \left[ w_+ (\sigma_k^+ \sigma_{k+1}^- - \hat{n}_k) + w_-( \sigma_k^- \sigma_{k+1}^+ 
- \hat{n}_{k+1}) + (w_+ + w_-) \rme^{2s} \sigma_k^+ \sigma_{k+1}^+   \right] 
\ee
For definiteness we consider an even number of sites $L=2K$. 
As initial distribution we take a product measure with density $\rho$.

We consider the generating function
\bel{3-4-2}
Y_s(t) = \exval{\rme^{sA^{ann}(t)}}=\bra{s} \rme^{-H^a(s) t} \ket{\rho}, \quad Y^\ast_s = \lim_{t\to \infty} Y_s(t)
\ee
and define a generalized large deviation function
\bel{3-4-3}
g(s) := \lim_{t\to \infty} t^{\alpha} \lim_{L\to\infty} \frac{1}{L} (\ln{Y^\ast_s} - \ln{Y_s(t)} ) =: 
\lim_{t\to \infty} t^{\alpha} (h^\ast_s - h_s(t)).
\ee
Observe the order of limits in this expression. Taking the limit $t\to\infty$ first would result trivially in $g(s)=0$
since in a finite lattice the system reaches the trivial steady state with no particles in finite time.
The total number of particles annihilated until time $t$ is extensive in system size. This motivates the
factor $1/L$ in \eref{3-4-3}. On the other hand,
one expects the annihilation activity per site to decay in time. Thus one expects $\alpha > 0$ and some
stationary constant $Y^\ast_s$. In fact,
here we show that $\alpha=1/2$ and we also present the full functions $g(s)$ and $h^\ast_s$.

\begin{theo}
For annihilating random walks as defined in \eref{3-4-1} with product initial distribution with strictly positive density $\rho>0$
the constant $h^\ast_s$ defined in \eref{3-4-3} takes the value 
\bel{3-4-5a}
h^\ast_s = 1- \rho(1-\rme^{s}).
\ee
The generalized large deviation function $g(s)$ has a non-trivial limit for $\alpha=1/2$ and does not depend
on the hopping asymmetry $w_+-w_-$
nor on the initial density $\rho$. Its explicit form is given by
\bel{3-4-5b}
g(s) = \frac{1}{4\sqrt{\pi(w_++w_-)t}} \sum_{m=1}^\infty \frac{(1-\rme^{-2s})^m}{m^{3/2}} .
\ee
\end{theo}

{\it Proof:}
By definition $A^{ann}(t)$ is the number of particles annihilated until time time $t$, i.e., 
$A^{ann}(t) = N(0) - N(t)$. Hence $Y_s(t) = \exval{\rme^{sA^{ann}(t)}} = \exval{ \rme^{s(N(0) - N(t))} }$.
On the other hand,
\be
\exval{\rme^{s(N(0) - N(t))}} = \bra{s} \rme^{-s\hat{N}} \rme^{-Ht} \rme^{s\hat{N}} \ket{\rho}.
\ee
Next we observe $\rme^{s\hat{N}} \ket{\rho} =  a_L(s) \ket{\tilde{\rho}(s)}$
with $\tilde{\rho}(s) =\rme^{s} \rho / [1-\rho(1-\rme^{s})]$ and $a_L(s) = [1-\rho(1-\rme^{s})]^L$.
Below we shall drop the argument $s$ of $\tilde{\rho}$.
This yields
\be
Y_s(t) = (h^\ast_s)^L \exval{\rme^{-s N(t)}}_{\tilde{\rho}}
\ee
with the function $h^\ast_s$ \eref{3-4-5a}. 

It remains to consider
$g(s) = -  \lim_{t\to \infty} t^{\alpha} \lim_{L\to\infty} 1/L \ln{\exval{\rme^{-s N(t)}}_{\tilde{\rho}} }$.
%The computation of the generating function of the particle number $\exval{\rme^{-s N(t)}}_{\tilde{\rho}}$ is non-trivial.
Since the particle number parity $(-1)^N$ is conserved we decompose the Bernoulli product measure with density
${\tilde{\rho}}$
into its even and odd components
\bea
P^+(\bfn) & = & \frac{1}{Z^+}\tilde{\rho}^N(1-\tilde{\rho})^{L-N} \sum_{M=0}^{K} \delta_{N,2M} \\
P^-(\bfn) & = & \frac{1}{Z^-}\tilde{\rho}^N(1-\tilde{\rho})^{L-N} \sum_{M=0}^{K-1} \delta_{N,2M+1}
\eea
where $N = \sum_{k=1}^L n(k)$ is the particle number in configuration $\bfn$ and
\be
Z^\pm=\frac{1}{2}\left(1\pm(1-2\tilde{\rho})^L\right).
\ee
The probability vectors
corresponding to these distributions are denoted by $\ket{\tilde{\rho}^{\pm}}$. 
We also define the projected summation vectors
\be
\bra{s^\pm} = \sum_{\{\bfn: (-1)^N = \pm 1\}} \bra{\bfn}.
\ee
This decomposition gives
\be
\exval{\rme^{-s N(t)}}_{\tilde{\rho}} = Z^+ \Phi_s^+(t) + Z^- \Phi_s^-(t)
\ee
with $\Phi_s^\pm(t) = \exval{\rme^{-s N(t)}}_{\tilde{\rho}^\pm}$.

The computation of $\Phi_s^+(t)$ was performed in \cite{Schu01}, using the mapping of annihilating random 
walks to a system of free
fermions \cite{Lush87}, and then employing a further mapping to $K=L/2$
non-interacting spins \cite{Schu95}. This allows for a tensor represention  of the even initial condition
on $(\C^2)^{\otimes K}$
\be
\ket{\tilde{\rho}^{+}} = \frac{1}{Z^+} \prod_{p=1}^{K} \left[ (1-\tilde{\rho})^2 + \tilde{\rho}^2 \cot{\frac{(2p-1)\pi}{2L}}\sigma_p^- \right] \ket{0},
\ee 
of the even summation vector 
\be
\bra{s^+} = \bra{0}  \prod_{p=1}^{K} \left[ (1+ \cot{\frac{(2p-1)\pi}{2L}}\sigma_p^+ \right] ,
\ee
and the even generator 
\be
H^+ = -(w_++w_-) \sum_{p=1}^{K}  \left[  \sin{\frac{(2p-1)\pi}{2L}}\sigma_p^+ - \left(1-\sin{\frac{(2p-1)\pi}{2L}}\right) \hat{n}_p \right].
\ee
Adapting the parameters in \cite{Schu01} with parameters $D$ and $\alpha$ 
to the present choice of parameters leads to the identification $s=-i\alpha$ and $D=(w_++w_-)/2$.
This yields
\be
\Phi_s^+(t) = \prod_{p=1}^{K} \left[ 1-\gamma^+_p(1-\rme^{-2s})\rme^{-2\epsilon^+_pt}\right]
\ee
with
\be 
\gamma^+_p = \frac{\tilde{\rho}^2}{\tilde{\rho}^2 + (1-\tilde{\rho})^2 \tan^2{\frac{(2p-1)\pi}{2L}}}, \quad
\epsilon^+_p = 2 D \left( 1- \cos{\frac{(2p-1)\pi}{L}} \right).
\ee

For the odd sector which was not studied previously we follow the approach of \cite{Schu95} to obtain
\be
\ket{\tilde{\rho}^{-}} = \frac{L\tilde{\rho}(1-\tilde{\rho})}{Z^-} 
\prod_{p=1}^{K-1} \left[ (1-\tilde{\rho})^2 + \tilde{\rho}^2 \cot{\frac{p\pi}{L}}\sigma_p^- \right] \sigma_0^-\ket{0}
\ee 
and the odd summation vector 
\be
\bra{s^-} = \bra{0} \sigma_0^+ \prod_{p=1}^{K-1} \left[ (1+ \cot{\frac{p\pi}{L}}\sigma_p^+ \right] .
\ee
The required normalization $\inprod{s^-}{\tilde{\rho}^{-}}=1$ follows from applying the trigonometric product identities
6.1.1.7. and 6.1.1.8. in \cite{Prud86}.
The generator on the odd subspace is similar to even case, but with the replacement $p \to p+1/2$. This yields
\be
\Phi_s^-(t) = \rme^{-s} \prod_{p=1}^{K} \left[ 1-\gamma^-_p(1-\rme^{-2s})\rme^{-2\epsilon^-_pt}\right]
\ee
with $\gamma^-_p$, $\epsilon^-_p$ similar to the corresponding even quantities, but with the replacement $p \to p+1/2$.
The exponential prefactor comes from the term involving $B^+_0$.

In order to take the thermodynamic limit one expresses the product as the exponential of the sum of logarithms. Since
all terms are well-behaved in the domain of finite $s$ and sufficiently large $t$ one can expand the logarithm in a power series.
Then using Euler-MacLaurin one can replace the sums by integrals and realizes that the even and odd product
converge to the same quantity.
Taking the thermodynamic limit thus yields $g(s)$ as given in \eref{3-4-5b}. 
\qed

Finally we consider a mixed process where particles coalesce with probability $p$ or annihilate with probability
$1-p$ when they meet on the same site.  The weighted generator for the integrated number of annihilated particles
is given by $H^{ac}(s) = H_0 + H^{ac}_1(s)$ with annihilation part
\bel{3-4-6}
H^{ac}_1(s) = - \sum_{k=1}^L \left[  (w_+ + w_-)(1-p) \rme^{2s} \sigma_k^+ \sigma_{k+1}^+ 
+ p w_+ \rme^{s} \sigma_k^+ \hat{n}_{k+1} + p w_- \rme^{s} \hat{n}_k^+ \sigma_{k+1}^+  \right] .
\ee
In the limit $p=1$ one has coalescing particles \cite{Arra81,benA98}. 
We define 
\bel{3-4-7}
Y^{ac}_s(t) := \rme^{sA^{ac}(t)}=\bra{s} \rme^{-H^{ac}(s) t} \ket{\rho}
\ee
and in complete analogy to \eref{3-4-3}
\bel{3-4-8}
g^{ac}(s) := \lim_{t\to \infty} t^{\alpha} \lim_{L\to\infty} \frac{1}{L} (\ln{Y^{\ast}_s} - \ln{Y^{ac}_s(t)} ) =: 
\lim_{t\to \infty} t^{\alpha} (h^{\ast}_s - h^{ac}_s(t)).
\ee
Here trivially the same function $Y^{\ast}_s$ appears since the initial measure is the same and
$Y^{\ast}_s$ is just the asymptotic generating function for the annihilation of all particles.

It was shown in \cite{Henk95}
that $H^{ac}(0) = B H^a(0) B^{-1}$ with $B=\rme^{-\theta(S^+-\hat{N})}$ and $\rme^{-\theta}=1-p/2$.
These transformations lead to $g^{ac}(s) = g(s')$ with some $s'(s,p)$. 
The lack of dependence of the
generalized large deviation function $g(\cdot)$ on the hopping asymmetry is trivial in the sense that is well-known that
for annihilating random walks the bias can be absorbed in a Galilei-transformation \cite{Schu95}. However,
there is a non-trivial universality property in the sense that $g(\cdot)$ does not depend on the
initial density and on the branching ratio.  We conjecture that the same generalized large deviation function will be found
in other models such as 
the SSEP with pair annihilation with arbitrary rate, which is also an integrable model \cite{Alca93,Schu01}.

\section*{Acknowledgements}
The author thanks B. Meerson and P.L. Krapivsky for inspiring discussions and 
the Galileo Galilei Institute for Theoretical Physics for hospitality and 
the INFN for partial support during the completion of this work.

\end{document}